\def\year{2020}\relax
\documentclass[letterpaper]{article} 
\usepackage{aaai20}  
\usepackage{times}  
\usepackage{helvet} 
\usepackage{courier}  
\usepackage[hyphens]{url}  
\usepackage{graphicx} 
\usepackage{amssymb}
\usepackage{graphicx}  
\title{Scalable and Generalizable Social Bot Detection through Data Selection}

\author{
Kai-Cheng Yang,\textsuperscript{\rm 1}
Onur Varol,\textsuperscript{\rm 2}
Pik-Mai Hui,\textsuperscript{\rm 1}
Filippo Menczer\textsuperscript{\rm 1,3} \\
\textsuperscript{\rm 1}Center for Complex Networks and Systems Research, Indiana University, Bloomington, IN, USA \\
\textsuperscript{\rm 2}Center for Complex Networks Research, Northeastern University, Boston, MA, USA \\
\textsuperscript{\rm 3}Indiana University Network Science Institute, Bloomington, IN, USA
}

\begin{document}

\maketitle

\begin{abstract}
Efficient and reliable social bot classification is crucial for detecting information manipulation on social media.
Despite rapid development, state-of-the-art bot detection models still face generalization and scalability challenges, which greatly limit their applications.
In this paper we propose a framework that uses minimal account metadata, enabling efficient analysis that scales up to handle the full stream of public tweets of Twitter in real time.
To ensure model accuracy, we build a rich collection of labeled datasets for training and validation.
We deploy a strict validation system so that model performance on unseen datasets is also optimized, in addition to traditional cross-validation.
We find that strategically selecting a subset of training data yields better model accuracy and generalization than exhaustively training on all available data. 
Thanks to the simplicity of the proposed model, its logic can be interpreted to provide insights into social bot characteristics. 
\end{abstract}

\section{Introduction}

Social bots are social media accounts controlled in part by software. 
Malicious bots serve as important instruments for orchestrated, large-scale opinion manipulation campaigns on social media \cite{ferrara2016rise,subrahmanian2016darpa}. Bots have been actively involved in online discussions of important events, including recent elections in the US and Europe \cite{bessi2016social,deb2019perils,stella2018bots,ferrara2017disinformation}.
Bots are also responsible for spreading low-credibility information \cite{shao2018spread} and extreme ideology \cite{berger2015isis,ferrara2016predicting}, as well as adding confusion to the online debate about vaccines \cite{broniatowski2018weaponized}.

Scalable and reliable bot detection methods are needed to estimate the influence of social bots and develop effective countermeasures. The task is
challenging due to the diverse and dynamic behaviors of social bots. For example, some bots act autonomously with minimal human intervention, others are manually controlled so that a single entity can create the appearance of multiple human accounts \cite{grimme2017social}. And while some bots are active continuously, others focus bursts of activity on different short-term targets. 
It is extremely difficult for humans with limited access to social media data to recognize bots, making people vulnerable to manipulation.

Many machine learning frameworks for bot detection on Twitter have been proposed in the past several years (see a recent survey \cite{alothali2018detecting} and the Related Work section). 
However, two major challenges remain: scalability and generalization.

\textit{Scalability} enables analysis of streaming data with limited computing resources. Yet most current methods attempt to maximize accuracy by inspecting rich contextual information about an account's actions and social connections. Twitter API rate limits and algorithmic complexity make it impossible to scale up to real-time detection of manipulation campaigns \cite{stella2018bots}.

\textit{Generalization} enables the detection of bots that are different from those in the training data, which is critical given the adversarial nature of the problem --- new bots are easily designed to evade existing detection systems. Methods considering limited behaviors, like retweeting and temporal patterns, can only identify certain types of bots. Even more general-purpose systems are trained on labeled datasets that are sparse, noisy, and not representative enough. As a result, methods with excellent cross-validation accuracy suffer dramatic drops in performance when tested on cross-domain accounts \cite{de2018lobo}.

In this paper, we propose a framework to address these two challenges. 
By focusing just on user profile information, which can be easily accessed in bulk, scalability is greatly improved.
While the use of fewer features may entail a small compromise in individual accuracy, we gain the ability to analyze a large-volume stream of accounts in real time. 

We build a rich collection of labeled data by compiling all labeled datasets available in the literature and three newly-produced ones. We systematically analyze the feature space of different datasets and the relationship between them; some of the datasets tend to overlap with each other while others have contradicting patterns.

The diversity of the compiled datasets allows us to build a more robust classifier. Surprisingly, by carefully selecting a subset of the training data instead of merging all of the datasets, we achieve better performance in terms of cross-validation, cross-domain generalization, and consistency with a widely-used reference system. We also show that the resulting bot detector is more interpretable. The lessons learned from the proposed data selection approach can be generalized to other adversarial problems.

\section{Related Work}

Most bot detection methods are based on supervised machine learning and involve manually labeled data \cite{ferrara2016rise,alothali2018detecting}.
Popular approaches leverage user, temporal, content, and social network features with  random forest classifiers \cite{davis2016botornot,varol2017online,gilani2017classification,yang2019arming}.
Faster classification can be obtained using fewer features and logistic regression \cite{ferrara2017disinformation,stella2018bots}.
Some methods include information from tweet content in addition to the metadata to detect bots at the tweet level \cite{kudugunta2018deep}.
A simpler approach is to test the randomness of the screen name \cite{beskow2019its}.
The method presented here also adopts the supervised learning framework with random forest classifiers, attempting to achieve both the scalability of the faster methods and the accuracy of the feature-rich methods. We aim for greater generalization than all the models in the literature.

Another strain of bot detection methods goes beyond individual accounts to consider  collective behaviors.
Unsupervised learning methods are used to find improbable similarities among accounts.
No human labeled datasets are needed. 
Examples of this approach leverage similarity in post timelines \cite{chavoshi2016debot}, action sequences \cite{cresci2016dna,cresci2018social}, content \cite{chen2018unsupervised}, and friends/followers \cite{jiang2016catching}.
Coordinated retweeting behavior has also been used \cite{mazza2019rtbust}.
Methods aimed at detecting coordination are very slow as they need to consider many pairs of accounts. The present approach is much faster, but considers individual accounts and so cannot detect coordination. 

\section{Feature Engineering}

All Twitter bot detection methods need to query data before performing any evaluation, so they are bounded by API limits.
Take Botometer, a popular bot detection tool, as an example.
The classifier uses over 1,000 features from each account \cite{varol2017online,yang2019arming}.
To extract these features, the classifier requires the account's most recent 200 tweets and recent mentions from other users.
The API call has a limit of 43,200 accounts per API key in each day.
Compared to the rate limit, the CPU and Internet i/o time is negligible.
Some other methods require the full timeline of accounts \cite{cresci2016dna} or the social network \cite{minnich2017botwalk}, taking even longer.

We can give up most of this contextual information in exchange for speed, and rely on just user metadata \cite{ferrara2017disinformation,stella2018bots}.
This metadata is contained in the so-called user object from the Twitter API. The rate limit for users lookup is 8.6M accounts per API key in each day. 
This is over 200 times the rate limit that bounds Botometer. 
Moreover, each tweet collected from Twitter has an embedded user object.
This brings two extra advantages.
First, once tweets are collected, no extra queries are needed for bot detection.
Second, while users lookups always report the most recent user profile, the user object embedded in each tweet reflects the user profile at the moment when the tweet is collected. This makes bot detection on archived historical data possible.

\begin{table*}
\centering
\caption{
List of features used in our framework.
User metadata features are extracted directly from the user object fetched from the API. Twitter stopped providing the geo\_enabled field as of May, 2019, so we remove it from the feature list although it proved helpful. 
Derived features are calculated based on the user metadata. Some are explained in the text, others are self-explanatory.
}
\begin{tabular}{l|l||l|l|l}
    \hline
    \multicolumn{2}{c||}{user metadata} & \multicolumn{3}{c}{derived features} \\
    \hline
    feature name & type & feature name & type & calculation\\
    \hline
    statuses\_count & count & tweet\_freq & real-valued & statuses\_count / user\_age \\
    followers\_count  & count & followers\_growth\_rate & real-valued &  followers\_count / user\_age \\
    friends\_count  & count & friends\_growth\_rate & real-valued & friends\_count / user\_age \\
    favourites\_count & count & favourites\_growth\_rate & real-valued & favourites\_count / user\_age \\
    listed\_count & count & listed\_growth\_rate & real-valued & listed\_count / user\_age \\
    default\_profile & binary & followers\_friends\_ratio & real-valued & followers\_count / friends\_count \\
    profile\_use\_background\_image & binary & screen\_name\_length & count & length of screen\_name string \\
    verified & binary & num\_digits\_in\_screen\_name & count & no. digits in screen\_name string \\
    & & name\_length & count & length of name string\\
    & & num\_digits\_in\_name & count & no. digits in name string\\
    & & description\_length & count & length of description string \\
    & & screen\_name\_likelihood & real-valued & likelihood of the screen\_name\\
    \hline
\end{tabular}
\label{table:features}
\end{table*}

Table~\ref{table:features} lists the features extracted from the user object.
The rate features build upon the user age, which requires the probe time to be available. When querying the users lookup API, the probe time is when the query happens. If the user object is extracted from a tweet, the probe time is the tweet creation time  (created\_at field).
The user\_age is defined as the hour difference between the probe time and the creation time of the user (created\_at field in the user object).
User ages are associated with the data collection time, an artifact irrelevant to bot behaviors.
In fact, tests show that including age in the model deteriorates accuracy. However, age is used to calculate the rate features.
Every count feature has a corresponding rate feature to capture how fast the account is tweeting, gaining followers, and so on.
In the calculation of the ratio between followers and friends, the denominator is  $\max(\mbox{friends\_count}, 1)$ to avoid division-by-zero errors.

The screen\_name\_likelihood feature is inspired by the observation that bots sometimes have a random string as  screen name \cite{beskow2019its}. Twitter only allows letters (upper and lower case), digits, and underscores in the screen\_name field, with a 15-character limit.
We collected over 2M unique screen names and constructed the likelihood of all 3,969 possible bigrams. 
The likelihood of a screen name is defined by the geometric-mean likelihood of all bigrams in it.
We do not consider longer n-grams as they require more resources with limited advantages.
Tests show the likelihood feature can effectively distinguish random strings from authentic screen names.

\section{Datasets}
\label{sec:datasets}

To train and test our model, we collect all public datasets of labeled human and bot accounts and create three new ones, all available in the bot repository (\url{botometer.org/bot-repository}).
See overview in Table~\ref{table:dataset}.

Let us briefly summarize how the datasets were collected.
The \textbf{\texttt{caverlee}} dataset consists of bot accounts lured by honeypot accounts and verified human accounts \cite{lee2011seven}. This dataset is older than the others.
To obtain the \textbf{\texttt{varol-icwsm}} dataset, the authors manually labeled accounts sampled from different Botometer score deciles \cite{varol2017online}.
The dataset was designed to be representative of diverse types of accounts.
The bot accounts in the \textbf{\texttt{cresci-17}} dataset contain a more fine-grained classification: traditional spambots, social spambots, and fake followers \cite{cresci2017paradigm}.
Traditional spambots are simple bots that tweet the same content repeatedly. 
Social spambots mimic the profiles and actions of normal users so they are not suspicious when inspected individually.
But the authors found them promoting certain hashtags or content in a coordinated fashion.
Fake followers are accounts paid to follow other accounts.
The \textbf{\texttt{pronbots}} dataset consists of a group of bots that share scam sites.
The dataset was first shared by Andy Patel (\url{github.com/r0zetta/pronbot2}) and then collected for study \cite{yang2019arming}. 
The \textbf{\texttt{celebrity}} dataset is made of accounts selected among celebrities \cite{yang2019arming}.
Accounts in the \textbf{\texttt{vendor-purchased}} dataset are fake followers purchased by researchers from several companies \cite{yang2019arming}.
The \textbf{\texttt{botometer-feedback}} dataset was constructed by mannually labeling accounts flagged by feedback from Botometer users \cite{yang2019arming}.
The \textbf{\texttt{political-bots}} dataset is a group of politics-oriented bots shared by Twitter user \texttt{@josh\_emerson}  \cite{yang2019arming}.
For the \textbf{\texttt{gilani-17}} dataset, accounts collected  using the Twitter streaming API were grouped into four categories based on the number of followers \cite{gilani2017bots}.
The authors then sampled accounts from the four categories and had four undergraduate students annotate them based on key information compiled in a table.
For the \textbf{\texttt{cresci-rtbust}} dataset, the authors collected all Italian retweets between 17--30 June 2018, then manually annotated a roughly balanced set of about 1,000 human and bot accounts \cite{mazza2019rtbust}.
Bots in the \textbf{\texttt{cresci-stock}} dataset were isolated by finding accounts with similar timelines among tweets containing selected cashtags collected during five months in 2017 \cite{cresci2018fake,cresci2019cashtag}.

We created three more datasets.
The \textbf{\texttt{midterm-18}} dataset was filtered based on political tweets collected during the 2018 U.S. midterm elections \cite{yang2019bot}.
We manually identified some of the genuine human users that were actively involved in the online discussion about the elections.
The bot accounts were spotted through suspicious correlations in their creation and tweeting timestamps.
Most of the bot accounts have been suspended by Twitter after the elections, which confirms our labeling.
The \textbf{\texttt{botwiki}} dataset is based on the \url{botwiki.org} archive of self-identified bot accounts. We manually removed inactive accounts and those from platforms other than Twitter.
Finally, the \textbf{\texttt{verified}} dataset was generated by filtering the streaming API for verified accounts.
This dataset is added as a supplement to balance \texttt{vendor-purchased} and \texttt{botwiki}, because the analyses in the following sections require the presence of both human and bot accounts. However, since the verified account flag is a feature of the classifier, we set this feature to `false' for the human accounts; this is a conservative choice that prevents any bias in favor of our model. 

By the time we collected the datasets, some of the accounts had been suspended already, so the numbers shown in Table~\ref{table:dataset} might be smaller than those in the original papers.

\begin{table}
\centering
   \caption{
   Datasets of labeled bot and human accounts.
   }
\begin{tabular}{l|r|r}
   \hline
   Dataset & \#bots & \#human\\
   \hline
   \texttt{caverlee} & 15,483 & 14,833\\
   \texttt{varol-icwsm} & 733 & 1,495\\
   \texttt{cresci-17} & 7,049 & 2,764\\
   \texttt{pronbots} & 17,882 & 0\\
   \texttt{celebrity} & 0 & 5,918\\
   \texttt{vendor-purchased} & 1,087 & 0 \\
   \texttt{botometer-feedback} & 139 & 380 \\
   \texttt{political-bots} & 62 & 0 \\
   \texttt{gilani-17} & 1,090 & 1,413\\
   \texttt{cresci-rtbust} & 353 & 340\\
   \texttt{cresci-stock} & 7,102 & 6,174\\
   \texttt{midterm-18}  & 42,446 & 8,092\\
   \texttt{botwiki} & 698 & 0\\
   \texttt{verified} & 0 & 1,987\\
   \hline
   Total & 94,124 & 43,396\\
   \hline
\end{tabular}
   \label{table:dataset} 
\end{table}

Beside the labeled datasets that served as ground truth, we also created a dataset of 100,000 random users collected from the streaming API in 2018.
This data was used for model evaluation, as discussed in a later section.

\section{Data Characterization}
\label{sec:data_characterization}

Since there are drastically different types of bot (and human) accounts \cite{de2018lobo}, characterizing the accounts in different datasets can provide us with useful insight into the design of a generalizable classifier.

\subsection{Independent Dataset Analysis}

The most intuitive way to inspect the different training datasets is to visualize them in feature space.
To highlight the contrast between human and bot accounts, we merge some of the single-class datasets in Table~\ref{table:dataset}.
Specifically, \texttt{pronbots} and \texttt{celebrity} are combined to form the \texttt{pron-celebrity} dataset; \texttt{botometer- feedback} and \texttt{political-bots} are merged into \texttt{political-feedback}; \texttt{verified} is split into two parts, merged with \texttt{botwiki} and \texttt{vendor-purchased} to obtain the roughly balanced  \texttt{bot\-wiki-verified} and \texttt{vendor-verified}, respectively.
The merging here is purely for analysis; we stick to the original datasets for data selection purposes below.

\begin{figure*}
    \centering
    \includegraphics[width=\textwidth]{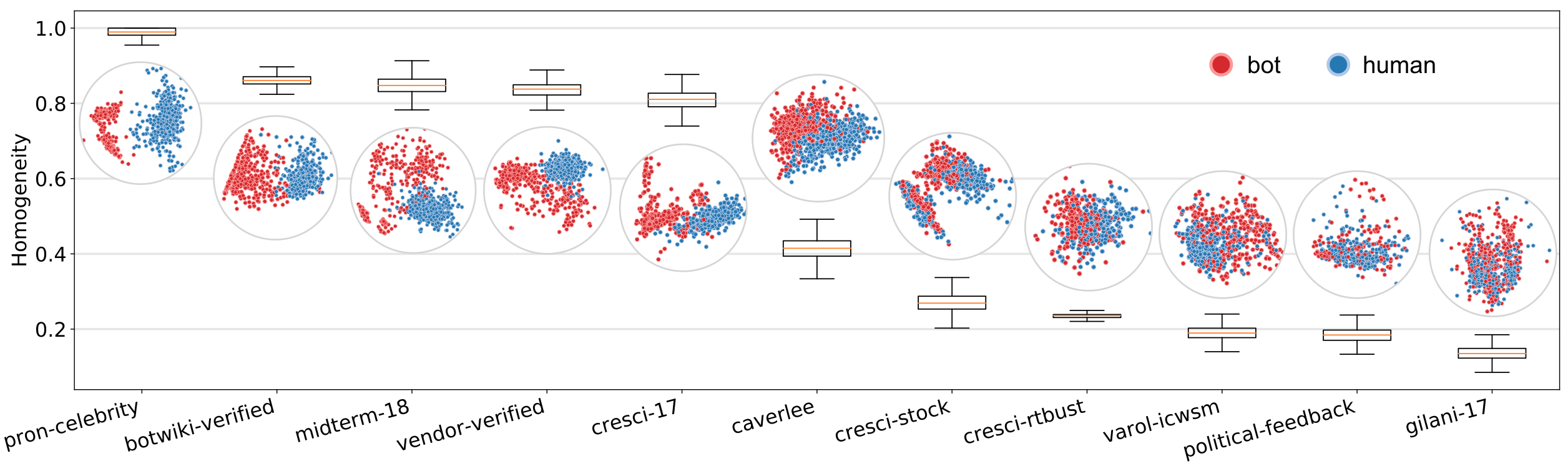}
    \caption{
    Visualization of human and bot accounts in different training datasets.
    The scatter plots visualize samples of 1,000 bot and human accounts after PCA.
    The box plots show the distributions of homogeneity scores across many samples from each dataset (see text for details).
    }
    \label{fig:accounts_visulization}
\end{figure*}

We apply PCA to project each dataset into the 2-D plane for visulization; t-SNE yields similar separation patterns. 
Since most of the features have wide ranges and skewed distributions, we first rescale them via log-transforms.
To quantify the separation of human and bot accounts in each dataset, we apply a kNN classifier in the original feature space.
For each account, we identify their nearest $k$ neighbors and assign the majority label to the focal account.
With the labels obtained from kNN and the ground truth, we are able to calculate the homogeneity score for each dataset \cite{rosenberg2007v}. 
The kNN algorithm is stable when $k>3$; we choose $k=9$ for our analysis. 
Because the datasets have various sizes and class ratios, we sample 500 bots and 500 humans (or fewer for datasets without enough accounts) to calculate the homogeneity score.
This procedure is repeated 1,000 times for each dataset to generate a distribution of the  scores. 

Fig.~\ref{fig:accounts_visulization} shows the PCA scatter plots and the homogeneity scores for all datasets. 
Five out of 11 datasets demonstrate a clearly clustered structure, suggesting bots and humans are easily separable using our features.
The rest of the datasets have clusters that are not as easily separable.
This is consistent with prior results showing no clear separation using t-SNE plots \cite{varol2017online,de2018lobo}.

Among the five less separable datasets, \texttt{cresci-stock} was labeled based on the timeline similarity between accounts.
Such coordination cannot be captured by our feature-based approach \cite{cresci2017paradigm,yang2019arming}.
The other four datasets are manually annotated and include different types of accounts; many of them exhibit behaviors that even humans find difficult to distinguish.
For example, \texttt{varol-icwsm} has 75\% inter-annotator agreement.
The mixed structure in feature space is understandable in light of these characteristics.

\subsection{Cross-Dataset Analysis}

Let us explore whether human and bot classes are consistent across different datasets.
Visualizing all the data together is unfeasible because we have too many datasets with too many data points. Let us instead use generalization as a proxy for consistency, by training the models on one dataset and testing on another. If the datasets are consistent, we expect good generalization power.

\begin{figure}[t!]
    \centering
    \includegraphics[width=\columnwidth]{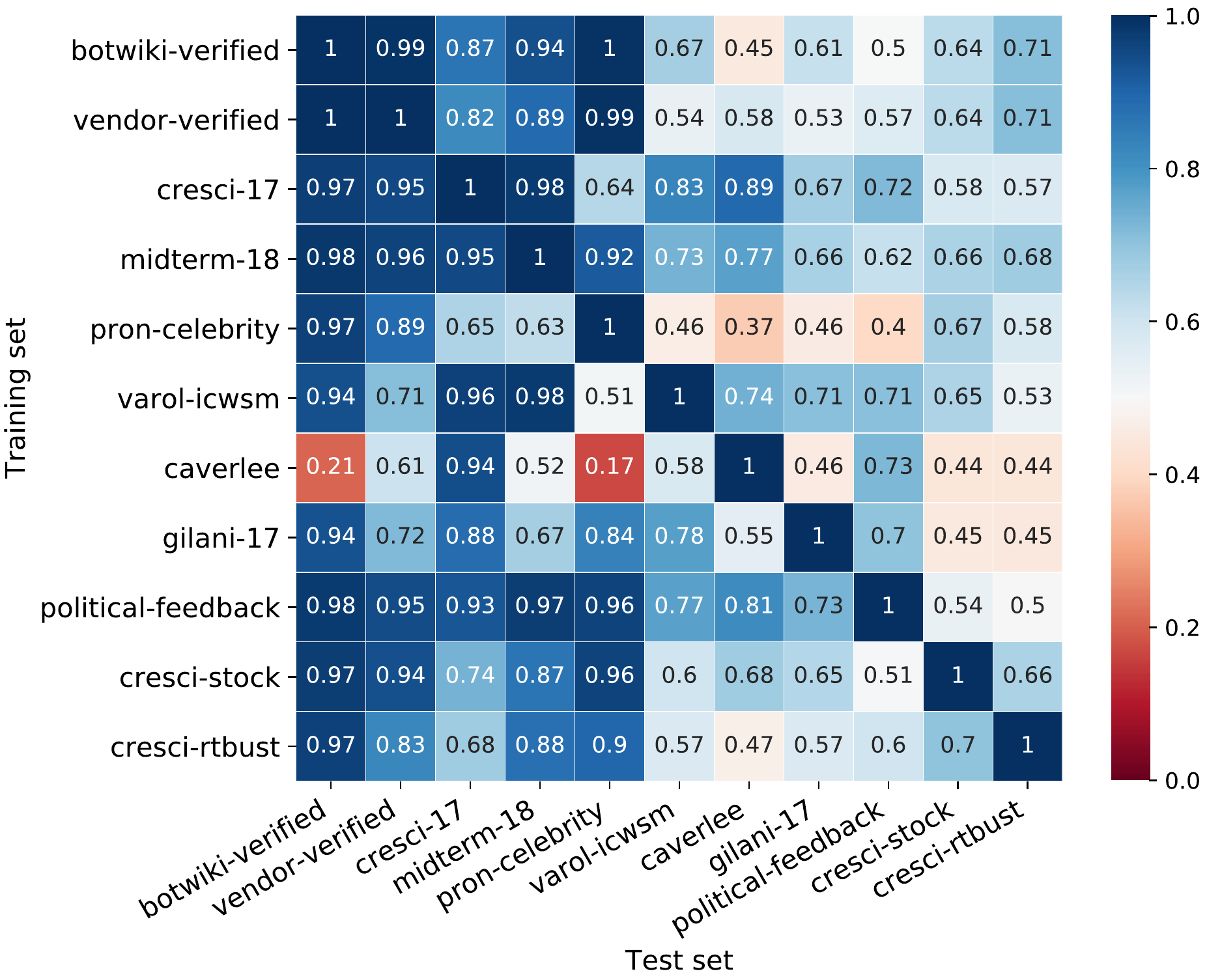}
    \caption{
    AUC scores of random forest classifiers trained on one dataset and tested on another.
    The datasets are ordered by separability (defined in the text).
    }
    \label{fig:dataset_matrix}
\end{figure}

The matrix in Fig.~\ref{fig:dataset_matrix} maps cross-dataset AUC using random forest classifiers. 
When tested on cross-domain data, the generalization AUC varies.
In some cases, it is very high indicating datasets with consistent classes.
But no dataset can generalize well on all other datasets.
In fact, the AUC can sometimes be below 0.5, suggesting that training and test datasets have contradictory labels.
Many factors may contribute to this phenomenon.
First, different datasets were annotated by different people with different standards using different methods.
Considering the difficult nature of the problem, labeling noise is not surprising.
Second, while our 20 features enable scalability, they only capture a tiny portion of an account's characteristics.
The seemingly contradicting patterns in our feature space might be resolved by considering additional features.

Another important observation is the asymmetry of Fig.~\ref{fig:dataset_matrix}: for a pair of datasets, exchanging the training and test sets may lead to different results.
We can define the 
\textit{separability} of a dataset by using it to test models trained on other classifiers --- averaging AUC across columns of the matrix in Fig.~\ref{fig:dataset_matrix}. For example, classifiers trained on most of the datasets achieve good performance on \texttt{botwiki-verified}, suggesting easy separability. On the other hand, \texttt{cresci-rtbust} is not easily separable because classifiers trained on other datasets perform poorly on it.
Similarly, we can define the 
\textit{generalizability} of a dataset by training a model on it and testing on other datasets --- averaging AUC across rows of the matrix in Fig.~\ref{fig:dataset_matrix}.
We find no clear correlation  between separability and generalizability (Spearman's $r=0.18$, $p=0.6$): the fact that bots in one dataset can be easily detected does not imply that a classifier trained on that dataset can detect bots in other datasets.

\section{Generalizability}

Random forest achieves perfect AUC when trained and tested on any single dataset, and excellent AUC in cross-validation \cite{alothali2018detecting,yang2019arming}; it is expressive enough to capture the non-linear patterns discriminating human and bot accounts even in the least separable datasets. 
Therefore, the proposed framework does not aim to provide a more expressive algorithm; we will stick with random forest.
Instead, our goal is to address the poor cross-dataset generalization highlighted in the previous section and in the literature \cite{de2018lobo}. 

\subsection{Model Evaluation}

We have seen that good cross-validation performance, even on multiple datasets, does not guarantee generalization across unseen datasets. We propose a more strict evaluation system where, in addition to cross-validation on training data, we set some datasets aside for cross-domain validation.
Those holdout datasets will act as unseen accounts for selecting models with best generalizability and give us a sense of how well the models perform when facing novel types of behavior.
Specifically, we use the datasets listed in Table~\ref{table:model_composition}, on which Botometer was trained \cite{yang2019arming}, as our candidate training datasets.
Botometer therefore serves as a baseline.
The rest of the datasets are holdout accounts.
The \texttt{cresci-stock} dataset is excluded from this experiment because the coordinated nature of the bots makes it unsuitable for training feature-based methods.
In addition to cross-validation and cross-domain validation, we also want to evaluate generalization to a broader, more representative set of accounts. To this end we employ the 100,000 random accounts as described earlier. 
The random accounts lack labels, therefore we use Botometer as a reference, since it has been tested and adopted in many studies \cite{Vosoughi1146,shao2018spread}. 

\subsection{Data Selection}

Throwing all training data into the classifier should give us the best model according to learning theory, if the labels are correct and the accounts are independent and identically distributed in the feature space \cite{shalev2014understanding}.
Unfortunately, our experiments in Fig.~\ref{fig:dataset_matrix} show that those assumptions do not hold in our case.
This suggests that generalization might be improved by training on a selected subset of the data. 

Our goal is to find a set of training accounts that optimizes our three evaluation metrics: cross-validation accuracy on training data, generalization to unseen data, and consistency with a more feature-rich classifier on unlabeled data.
Similar data selection methods have proven effective in other domains with noisy and  contradictory data \cite{wu2007data,erdem2010ransac,zhang2014agreement}.

\begin{figure}[t!]
    \centering
    \includegraphics[width=\columnwidth]{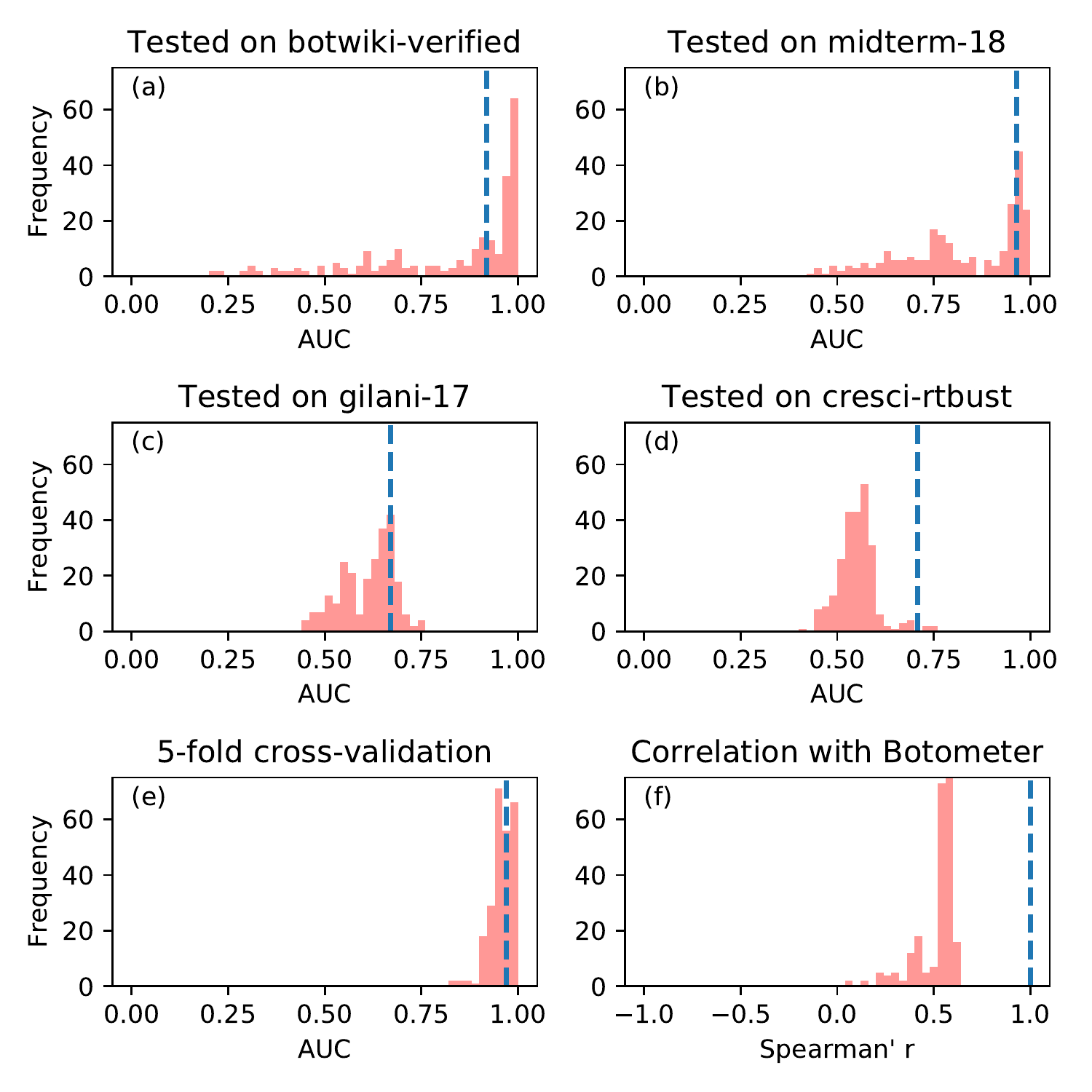}
    \caption{
    (a-d)~AUC distributions of candidate models tested on 4 unseen datasets. (e)~Five-fold cross-validation AUC distribution of candidate models. (f)~Distribution of Spearman's rank correlation coefficients between candidate models and Botometer on 100,000 random accounts.
    Baselines from Botometer are highlighted by dashed blue lines.
    }
    \label{fig:model_selection}
\end{figure}

We treat every dataset listed in Table~\ref{table:model_composition} as a unit and end up with 247 possible combinations with both human and bot accounts present.
Random forest classifiers with 100 trees are trained on those 247 combinations, yielding as many candidate models. We record the AUC of each model via five-fold cross-validation.
The classifiers are then applied to unseen datasets (see list in Table~\ref{table:performance_on_wild}) and random accounts.
Results are reported in Fig.~\ref{fig:model_selection}.
Cross-validation yields high AUC for most of the models, as expected.
For unseen accounts, datasets with high homogeneity tend to yield better AUC, whereas performance is worse on datasets with lower homogeneity.
Most of the models have positive correlation with Botometer on random accounts.
Botometer performs well in all AUC tests, although some candidate models beat Botometer in each test. No candidate model, however, outperforms the baseline on all unseen datasets.

To select a model that performs consistently well in all tests, we first rank the models in each of the six tests shown in Fig.~\ref{fig:model_selection} based on their performance, with the top model first. We then select the model with minimal product of the six ranks.
Models selected in this way may not achieve the best results in every single test, but will do well in all tests, ensuring stability in applications.
The datasets selected by the top 3 models are shown in Table~\ref{table:model_composition}; \texttt{M196} is the best model. Detailed performance metrics are reported in Table~\ref{table:performance_on_wild}. 

\begin{table}
\centering
  \caption{
   Datasets used to train selected candidate models and Botometer.
   \texttt{M196}, \texttt{M125} and \texttt{M191} are the top three models according to our selection method.
   }
\resizebox{.95\columnwidth}{!}{
\begin{tabular}{l|c|c|c|c|c}
   \hline
   Dataset                     &Botometer & M196 & M125 & M191 & M246 \\
   \hline
   \texttt{caverlee}           &\checkmark&          &          &          &\checkmark\\
   \texttt{varol-icwsm}        &\checkmark&\checkmark&\checkmark&\checkmark&\checkmark\\
   \texttt{cresci-17}          &\checkmark&\checkmark&\checkmark&\checkmark&\checkmark\\
   \texttt{pronbots}           &\checkmark&          &          &          &\checkmark\\
   \texttt{celebrity}          &\checkmark&\checkmark&\checkmark&\checkmark&\checkmark\\
   \texttt{vendor-purchased}   &\checkmark&          &          &\checkmark&\checkmark\\
   \texttt{botometer-feedback} &\checkmark&\checkmark&\checkmark&\checkmark&\checkmark\\
   \texttt{political-bots}     &\checkmark&\checkmark&          &          &\checkmark\\
   \hline
\end{tabular}
}
    \label{table:model_composition} 
\end{table}

\begin{table}
\centering
\caption{
    AUC scores on unseen datasets, five-fold cross-validation AUC, and correlation with Botometer for selected candidate models.
    Performance of the Botometer baseline and \texttt{M246} (trained on all data) is shown for comparison.
    Metrics significantly better than the baseline ($p<0.01$) are highlighted in bold.
}
\resizebox{.95\columnwidth}{!}{
\begin{tabular}{l|c|r|r|r|r}
\hline
Metric                        &Botometer&  M196 &     M125 &     M191  & M246\\
\hline
\texttt{botwiki-verified}  &0.92 & \textbf{0.99} & \textbf{0.99} & \textbf{0.99} & 0.91\\
\texttt{midterm-18}        &0.96 & \textbf{0.99} & \textbf{0.99} & \textbf{0.98} & 0.83\\
\texttt{gilani-17}         &0.67 & \textbf{0.68} & \textbf{0.69} & \textbf{0.68} & 0.64\\ 
\texttt{cresci-rtbust}     &0.71     & 0.60 &    0.59 &    0.58 & 0.57\\ \hline
5-fold cross-validation        &0.97     & \textbf{0.98} & \textbf{0.98} & \textbf{0.98} & 0.95\\ \hline
Spearman's r               &1.00     & 0.60 &    0.60 &    0.62 & 0.60\\
\hline
\end{tabular}
}
\label{table:performance_on_wild}
\end{table}

The winning models achieve very high AUC in cross-validation as well as on unseen datasets.
Further analysis reveals that the winning models all have much higher precision than the baseline in all holdout tests at the 0.5 threshold. 
Those models also achieve better recall exept for \texttt{gilani-17} and \texttt{cresci-rtbust}.
Accounts in these two datasets are annotated  with different types of behaviors.
In fact, when we train random forest models on these two datasets alone with five-fold cross-validation, we obtain AUC scores of 0.84 and 0.87 respectively, indicating challenging classification tasks. Botometer's generalization performance is also lower on these two datasets.
All things considered, the generalizability of the winning models seems acceptable.

\begin{figure}
    \centering
    \includegraphics[width=\columnwidth]{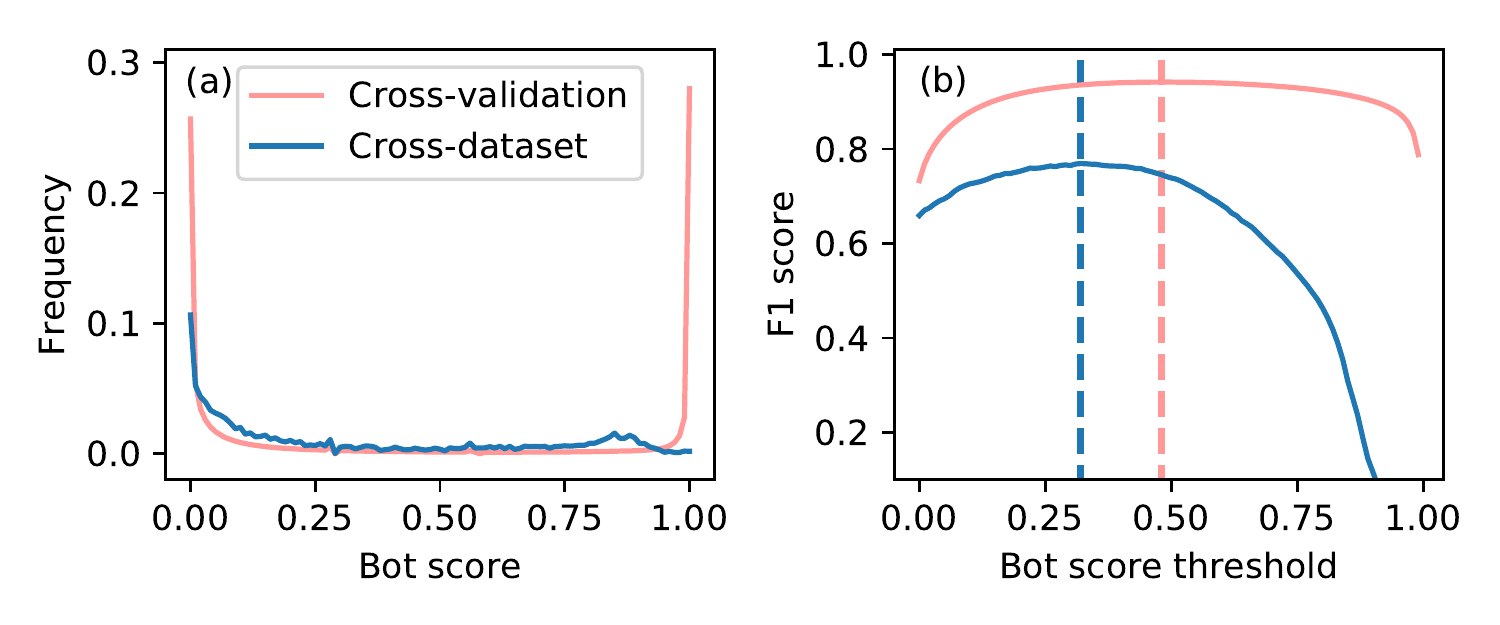}
    \caption{
    (a) Bot score distributions and (b) $F_1$ score versus bot score threshold, for the \texttt{M196} model. The best $F_1$ is obtained with thresholds of 0.48 and 0.32 for cross-validation and cross-domain testing, respectively (dashed lines).
    }
    \label{fig:threshold}
\end{figure}

Random forest generates a score between 0 and 1 to estimate the likelihood of an account exhibiting bot-like behavior. If we need a binary classifier, we can use a threshold. Fig.~\ref{fig:threshold} illustrates the thresholds that maximize precision and recall (via the $F_1$ metric). Different thresholds yield best accuracy in cross-validation ($F_1=0.94$, $R=0.93$, $P=0.94$) or cross-domain testing ($F_1=0.77$, $R=0.68$, $P=0.88$), underscoring that the choice of threshold depends on the holdout datasets used for validation. Depending on the topic of interest, practitioners can choose or create new datasets of annotated accounts for cross-domain validation and use the rest for training data selection.

\section{Scalability}

In this section, we quantify the scalability of the proposed framework.
The user metadata required for classification can be obtained through the users lookup endpoint or the streaming API, since every tweet carries the user object.
Users lookup allows checking 8.6M accounts per day with a user API key.
The maximum streaming volume is provided by the Firehose, which delivers all public tweets --- currently 500 millions per day on average \cite{fedoryszak2019real}.

We conducted an offline experiment to estimate the classification speed.
Our classifier was implemented in Python with scikit-learn \cite{scikit-learn} and run on a machine with an Intel Core i7-3770 CPU (3.40GHz) and 8GB RAM.
It takes an average of $9{,}612 \pm 6 \times 10^{-8}$ seconds to evaluate each tweet, which means almost 900M tweets per day, well beyond the Firehose volume.

\section{Model Interpretation}

\begin{figure}[!t]
    \centering
    \includegraphics[width=\columnwidth]{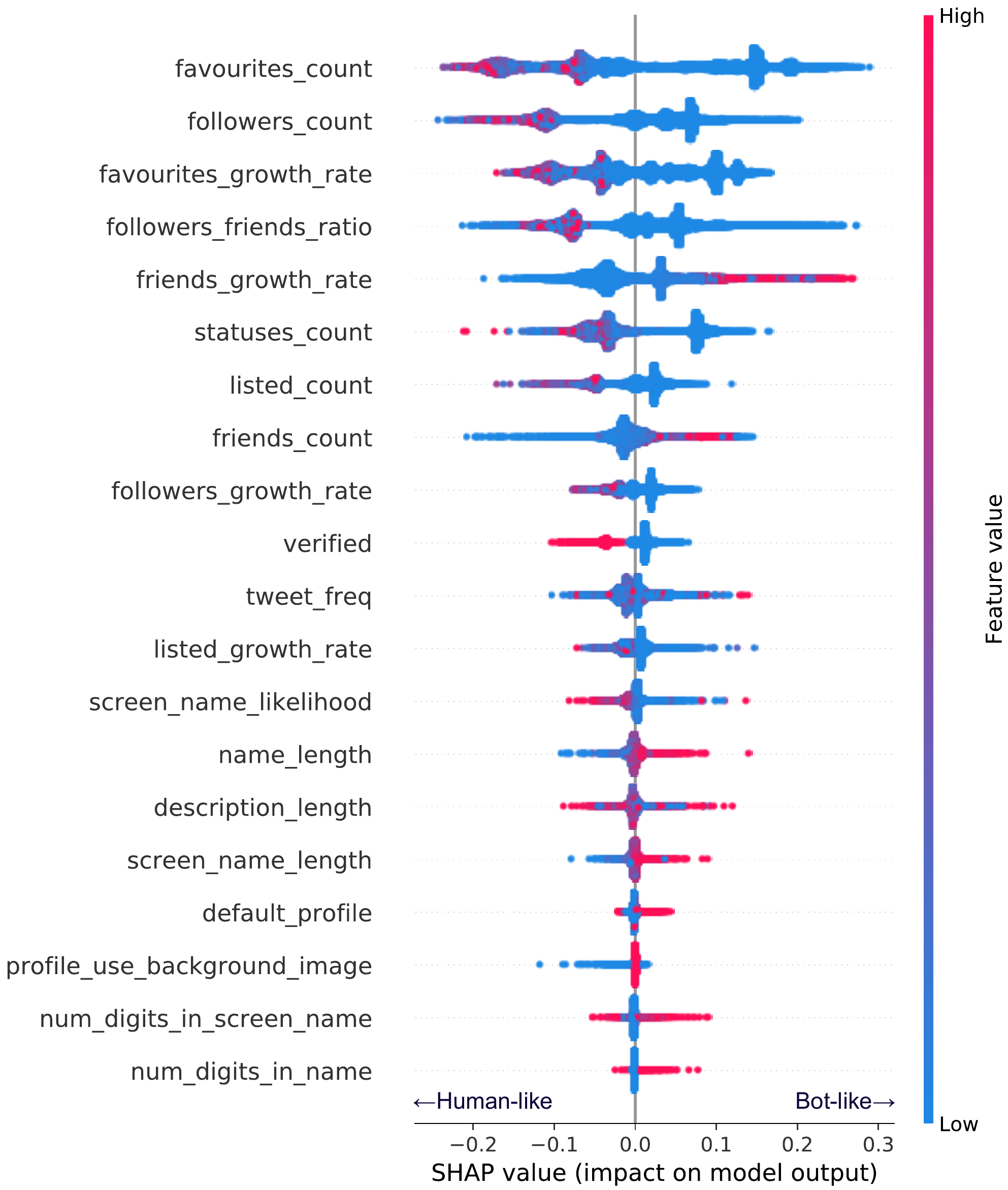}
    \caption{
    The summary plot generated by SHAP for model \texttt{M196}.
    The features are ordered by  importance from top to bottom.
    The x-axis shows the SHAP value of each feature.
    A positive SHAP value means the feature is pushing the result to 1 (bot-like), a negative SHAP value means the feature is pushing the result to 0 (human-like).
    The feature value is indicated by color: red means the feature value is large, blue means it's small. Binary features are coded so that true is represented by 1.
    }
    \label{fig:shap}
\end{figure}

With only 20 features in our best model (\texttt{M196}), it is possible to interpret its logic using the SHAP model explanation technique \cite{lundberg2018consistent}. 
Fig.~\ref{fig:shap} shows that, for example, long screen names (red) have positive SHAP values, meaning more bot-like. Conversely, verified accounts (red) have negative SHAP values, meaning more human-like.
The SHAP analysis tells us that high favorites and followers counts, favorites growth rate, and followers-friends ratio are signs of organic accounts.
High friends count and growth rate, however, suggest an account is suspicious.
This confirms the intuition that bots are eager to expand their friend lists, but do not attract as many followers.

Naturally, Fig.~\ref{fig:shap} presents an over-simplified picture of the effects of different features. 
The actual logic is more complex due to interactions among features.
For example, a low favourites count generally means bot-like, but if the account also has a high follower growth rate, then the effect of the favorites count is reduced.
Moreover, the interpretation depends on the specific data used to train the model.
Yet, this example illustrates an advantage of simplified feature sets.

\section{Conclusion}

We proposed a bot detection framework that scales up to real-time processing of the full public Twitter stream and that generalizes well to accounts outside the training data. 
Analysis of a rich collection of labeled data reveals differences in separability and generalization, providing insights into the complex relationship among datasets. 
Instead of training on all available data, we find that a subset of training data can yield the model that best balances performance on cross-validation, cross-domain generalization, and consistency with a widely adopted reference.
Thanks to the simplicity of the model, we are able to interpret the classifier.
The scalability opens up new possibilities for research involving analysis on massive social media data.
Our framework can also be embedded in other algorithms or systems for detecting more complex adversarial behaviors \cite{Hui2019BotSlayer}, such as coordinated influence campaigns, stock market manipulation, and amplification of misinformation.

Our work shows that data selection is a promising direction for dealing with noisy training data.
Developing smarter algorithms that can perform fine-grained data selection for better performance is an open challenge for future work.

\paragraph{Acknowledgments.}
We thank Emilio Ferrara for suggesting the \url{botwiki.org} archive. K.-C.~Y. and F.~M. were supported in part by Craig Newmark Philanthropies. P.-M.~H. and F.~M. were supported in part by DARPA contract W911NF-17-C-0094. 

\fontsize{9.5pt}{10.5pt}
\selectfont
\bibliography{ref}
\bibliographystyle{aaai}

\end{document}